\documentclass[conference]{IEEEtran}
\IEEEoverridecommandlockouts
\usepackage{cite}
\usepackage{amsmath,amssymb,amsfonts}
\usepackage{algorithmic}
\usepackage{graphicx}
\usepackage{subfigure}
\graphicspath{ {./plots/} }
\usepackage{textcomp}
\usepackage{xcolor}
\usepackage[T1]{fontenc}
\usepackage{xcolor, soul}
\usepackage{array}
\usepackage{multirow}
\usepackage{graphicx}
\usepackage{cuted}
\usepackage{lipsum}
\usepackage{algorithm}

 \newtheorem{lemma}{Lemma}
 \newtheorem{theorem}{Theorem}

 \newtheorem{definition}{Definition}

{}
{}

\makeatletter
\renewcommand*\env@matrix[1][\arraystretch]{%
  \edef\arraystretch{#1}%
  \hskip -\arraycolsep
  \let\@ifnextchar\new@ifnextchar
  \array{*\c@MaxMatrixCols c}}
\makeatother

\def\BibTeX{{\rm B\kern-.05em{\sc i\kern-.025em b}\kern-.08em
    T\kern-.1667em\lower.7ex\hbox{E}\kern-.125emX}}
\begin{document}

\title{Conditional Value-at-Risk for Quantitative Trading: A Direct Reinforcement Learning Approach}

\author{Ali~Al-Ameer,~\IEEEmembership{Member,~IEEE,}
        Khaled~Alshehri,~\IEEEmembership{ Member,~IEEE}
\thanks{A. Al-Ameer and K. Alshehri are with the Control and Instrumentation Engineering Department, College of Engineering and Physics, King Fahd University of Petroleum and Minerals (KFUPM), Dhahran, 31261, Saudi Arabia. Email: \{g201904230,kalshehri\}@kfupm.edu.sa. K. Alshehri is also with the Interdisciplinary Research Center for Smart Mobility and Logistics, KFUPM.}}

\maketitle

\begin{abstract}
    We propose a convex formulation for a trading system with the Conditional Value-at-Risk as a risk-adjusted performance measure under the notion of Direct Reinforcement Learning. Due to convexity, the proposed approach can uncover a lucrative trading policy in a "pure" online manner where it can interactively learn and update the policy without multi-epoch training and validation. We assess our proposed algorithm on a real financial market where it trades one of the largest US trust funds, SPDR, for three years. Numerical experiments demonstrate the algorithm's robustness in detecting central market-regime switching. Moreover, the results show the algorithm's effectiveness in extracting profitable policy while meeting an investor's risk preference under a conservative frictional market with a transaction cost of 0.15\% per trade.
\end{abstract}


\section{Introduction}
\par Financial markets exhibit severe stochasticity with non-stationary behavior \cite{cont2001empirical}. With this, the financial market can be viewed as a stochastic system with high uncertainties. To optimize a financial performance measure, that system, indeed, requires tools from control theory. Recent developments in control provided strong tools that are yet to be utilized in the field of quantitative finance, in particular asset allocation and portfolio management problems. Such problems can be looked at through the lens of stochastic optimal control \cite{davis1990portfolio}. 

The emergence of the Reinforcement Learning (RL) \cite{sutton2015RLintro}, a machine learning paradigm that intersects with optimal control theory, motivated researchers to explore its application in the field of asset allocation and portfolio optimization. The first two works that considered formulating the problem using RL were the ones by Neuneier \cite{neuneier1998enhancing}, and Moody and Saffel \cite{moody2001learning}. Neuneier implemented estimation of value functions with Q-learning while Moody and Saffel explored the use of direct policy search, which they called Direct Reinforcement Learning (DRL). Therefore, the adoption of RL as a tool to solve asset allocation and portfolio optimization problems is appealing since, under that paradigm, the controller can learn an optimal policy in an online manner.

\par Formulating the problem under the notion of DRL composes of several main parts. The first is the sensory part responsible for sensing the financial market's environment (system). The other part generates trading signals (control) following a parametric policy based on the environment's states\footnote{We will use the terms system and environment, controller and agent interchangeably in this paper.}. Lastly, we evaluate the policy pursued by the controller through a "reinforcing signal" that is received from the environment. That signal acts as feedback of the policy effectiveness in optimizing a predefined financial performance measure. Noteworthy, the mathematical properties of performance measures play a crucial role in determining the quality of the final obtained policy. That, in particular, motivates us to explore performance measures that can introduce convexity into the problem. Thus, we present the Conditional Value-at-Risk ($\mathrm{{\sf CVaR}}$) as the financial measure in this work. To the best of our knowledge, this is the first work that proposes a convex performance measure with $\mathrm{{\sf CVaR}}$ in the development of a trading system under the notion of DRL. 
\par The pioneering work of Rockafellar and Uryasev \cite{rockafellar2002conditional} was the first that introduced $\mathrm{{\sf CVaR}}$. We can fundamentally define $\mathrm{{\sf CVaR}}$ as the expected low-probable but significant losses that may incur during the investment horizon. Mathematically, $\mathrm{{\sf CVaR}}$ is a statistical quantity that measures the expectation of large losses that fall within the tail of the loss distribution. Due to the appealing mathematical properties of $\mathrm{{\sf CVaR}}$, such as convexity, it has grabbed the attention of researchers in the field of finance. In \cite{zhu2009worst,norton2021calculating,krokhmal2002portfolio}, the portfolio optimization problem was solved with $\mathrm{{\sf CVaR}}$ being an objective function or a constraint. In \cite{chow2017risk}, a risk-constrained RL algorithm that maximizes the expectation of the value function while meeting a constraint on {\sf CVaR} was proposed. Moreover, a DRL algorithm that searches for $\mathrm{{\sf CVaR}}$-optimal policy was proposed in \cite{keramati2020being}. 
\par In this paper, we propose a trading system that follows the DRL notion introduced in \cite{moody2001learning} while considering $\mathrm{{\sf CVaR}}$ as the system's performance measure. Our selection of $\mathrm{{\sf CVaR}}$ is motivated by financial and mathematical aspects. Financially, searching for an agent's policy based on optimizing $\mathrm{{\sf CVaR}}$ prevents the agent from making risky decisions that may cause significant losses. Therefore, it learns a policy that is aware of risk management under the severe uncertainties within the financial market. From the mathematical aspect, under careful formulation using $\mathrm{{\sf CVaR}}$, we can ensure convexity of the problem while keeping it general. The convex formulation provides provision for finding an optimal global policy, which allows for a fast online algorithm and leads to faster convergence. In fact, all of our experiments in this paper were implemented in a pure online manner, where the controller can interactively learn and update its policy "on the go" without multi-epoch training and validation. Moreover, they demonstrate the controller's robustness in handling the underlying uncertainties in the financial market. Noteworthy, convexity is not usual in quantitative finance, making our method appealing, avoiding solutions at local minimum that may cause unpredicted portfolio performance with high incurred losses \cite{sato2019model}. 
\subsection{Related Work}\label{sec:relWork}
To the best of our knowledge, Moody and Saffel were the first who introduced the concept of policy search RL into trading problems under the name of DRL \cite{moody2001learning}. They endeavored to design subtle performance measures that enhance the controller's risk awareness while trading. Their first measure was the Sharpe ratio, which is defined as the mean of achieved investment returns over its standard deviation \cite{sharpe1994sharpe}. In that sense, the Sharpe ratio is a symmetric risk-performance measure since it counts for the variance of both upside and downside returns. They, therefore, developed an asymmetric risk measure that counts for the downside returns, which they called the Downside Deviation ratio. Bertoluzzo and Corazza \cite{bertoluzzo2007making} explored a performance measure that maximizes the upside returns while minimizing the downside risk. In that sense, the learned policy is eager to maximize the positive returns while simultaneously avoiding decisions that may result in losses. In \cite{deng2015sparse, deng2016deep}, total returns and Sharpe ratio as the system's performance measures were implemented using a sparse-coding algorithm and deep fuzzy learning, respectively. Almahdi and Yang \cite{almahdi2019constrained} designed the performance function following the concept of Calmar ratio, which is defined as the ratio of expected returns to the expected maximum drawdown (MDD), i.e., the largest observed loss from a top of wealth to a trough \cite{magdon2004maximum}. Noteworthy, the two performance measures, total returns, and Sharpe ratio, are the most common among others and they were extensively scouted with different deep learning architectures \cite{lu2017agent,jiang2017cryptocurrency,weng2020portfolio,lei2020time}. Although most of the works report a promising result of their trading systems, their performance measures are non-convex. Hence, the optimization algorithms will probably converge to a local optimum, especially in deep learning algorithms where the problem is highly non-convex. To this end, we believe that our work takes the research of trading systems with the DRL notion to the next level, where we introduce $\mathrm{{\sf CVaR}}$ as a risk-performance measure that leads to a convex formulation for the problem. 
\subsection{Contributions}
In this work, our contributions into the field of developing trading systems under the notion of DRL are summarized as follows:
\begin{itemize}
    \item We bridge the gap between quantitative trading via DRL and the literature by introducing $\sf {\sf CVaR}$ as a performance measure.
    \item We provide sufficient conditions for the convexity of our model, which ensures superior performance for various setups and considerations.
    \item We propose an online algorithm that can actively discover and update a lucrative policy through a single run over the data. In other words, our proposed system can learn without multi-epoch training and validation.
\end{itemize}
\par The manuscript is organized as follows. We discuss the problem's preliminaries in Section II. Then, we provide our exposition for the proposed model in Section III. The numerical experiment is exhibited in Section IV. Section V discusses several potential generalizations for our proposed model, and the manuscript is concluded by Section VI. 
\section{Preliminaries}
\par Let $p_1, p_2, ..., p_t$ denote the price quotes of the financial asset, which are released by the exchange centers. We let the financial market represent the controller's environment, where its states are functions of the past $n$ prices, and are denoted by the vector $x_t:= h_t(p_t, p_{t-1}, ..., p_{t-n})$, where $h_t$ is continuous for each $t$. When the controller senses the environment's state $x_t$, it takes an action following a deterministic policy denoted by $\pi_t \in[-1,1]$, and any $\pi_t>0$ indicates a long position while $\pi_t<0$ indicates a short position\footnote{Long position is buying the asset with expecting the price to increase over time. The short position is borrowing the asset and selling it to a third party while expecting the price to drop over time. Later on, the asset is repurchased and returned to the lender in exchange for cash.}. Based on the action $\pi_t$ at a state $x_t$, the controller receives a reinforcing signal from the environment that represents the achieved return at the immediate next time step, denoted by $R_{t+1}$. The trading process of sensing the environment state, taking action and receiving a reinforcing signal repeats at each time step for $t=1, 2, 3, ..., T$ where $T$ is the investment horizon. 
\par Considering an investment's capital (or wealth) $W_0$ at time $t=0$, the controller's main objective is to learn a policy that optimizes a risk-adjusted performance measure over the investment horizon $T$, denoted by $\mathcal{U}_T(W_0,R)$ where $R \in \mathbb{R}^T$ is a reward vector representing received reinforcing signals over the investment horizon. In our work, for the performance measure, we are interested in utilizing the well-known risk measure $\mathrm{{\sf C{\sf VaR}}}$. At a probability level $\gamma$, $\mathrm{{\sf C{\sf VaR}}}_\gamma$ is the conditional mean of all losses that exceed the Value-at-Risk ($\mathrm{{\sf VaR}}$), where $\mathrm{{\sf VaR}}$ specifies the lowest loss $\beta$ that will not be exceeded  at a confidence level $\gamma$, and denoted by $\mathrm{{\sf VaR}}_\gamma$. We have the following definitions. 

\definition{
    Considering a continuous random loss denoted by $Z$ with a probability distribution function $\Psi_Z(z)$, then $\mathrm{{\sf VaR}}$ at a probability level $\gamma$ is given by:
    \begin{equation}
        \mathrm{{\sf VaR}}_\gamma(Z) = \min\{\beta \in \mathbb{R}: \Psi_Z(z)\geq\gamma\}
    \end{equation}}

\begin{figure}[!t]
     \centering
     \includegraphics[width=1\linewidth,bb= 0 0 375 230]{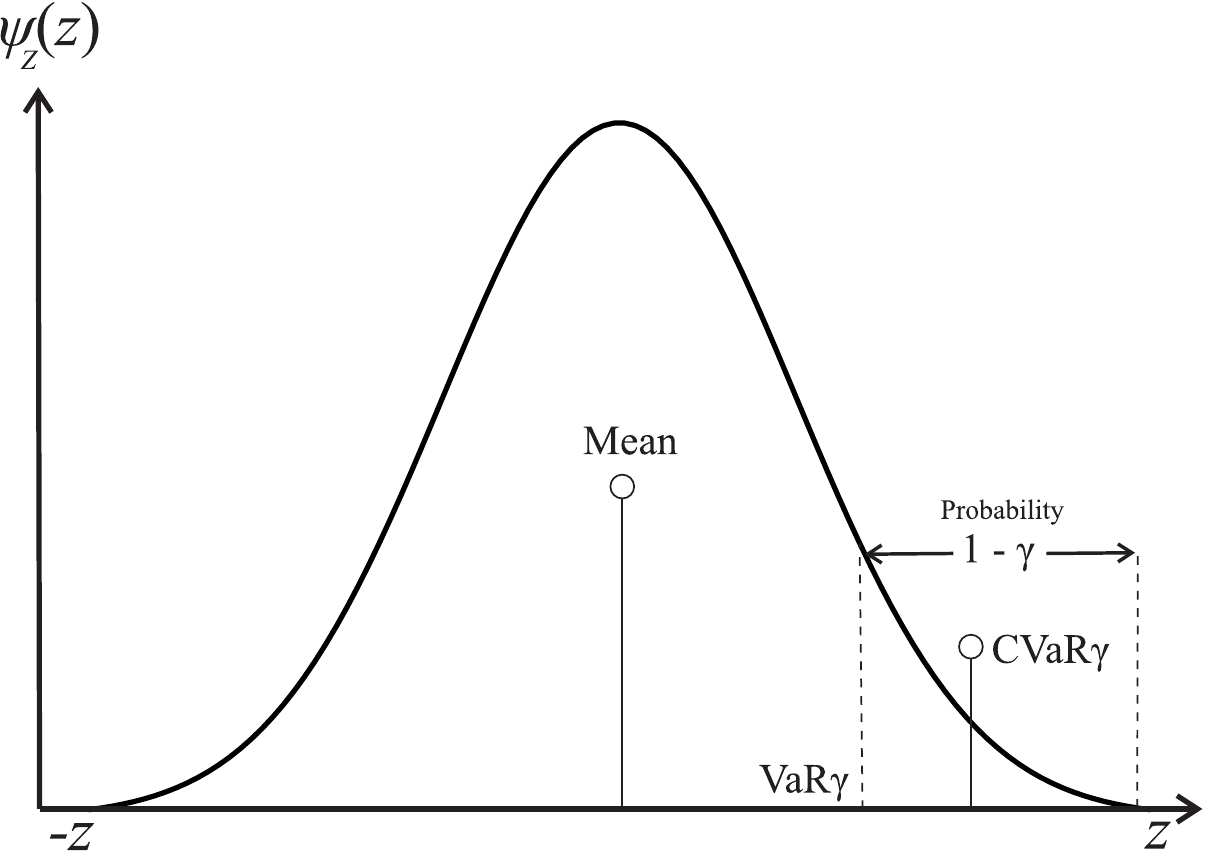}
     \caption{A general probability density for a loss with its respective $\mathrm{{\sf VaR}}_\gamma$ and $\mathrm{{\sf C{\sf VaR}}}_\gamma$}\label{fig:normalCVaR}
\end{figure}

\begin{definition}
    Considering a continuous random loss denoted by $Z$ with a probability density function $\psi_Z(z)$, then $\mathrm{{\sf C{\sf VaR}}}$ at a probability level $\gamma$ is given by:
    \begin{equation}
        \mathrm{{\sf C{\sf VaR}}}_\gamma(Z) = \frac{1}{1-\gamma} \int_{z\geq\mathrm{{\sf VaR}}_\gamma} z\psi_Z(z)dz
    \end{equation}
\end{definition}
Fig. \ref{fig:normalCVaR} illustrates a general example for $\psi_Z(z)$ with its $\mathrm{{\sf VaR}}_\gamma$ and $\mathrm{{\sf C{\sf VaR}}}_\gamma$. We emphasize that the confidence level $\gamma$ also plays the role or risk-aversion parameter where the controller is more risk-averse as $\gamma\rightarrow1$, and less risk-averse as $\gamma\rightarrow0$. With {\sf CVaR} as our risk-sensitive measure, we formulate our problem next.
\section{Proposed Model}
One can consider a general risk-sensitive problem that is to be optimized over an investment horizon of $t=1, 2, ..., T$:
\begin{equation}
    \begin{aligned}
        &\min\quad\mathcal{U}_T(W_0,R)\\ 
    \end{aligned}
    \label{eq:utility}
\end{equation} 
In our case, the selection of the risk-adjusted performance is motivated by the attractive mathematical properties found in ${\sf CVaR}$. We, therefore, are interested in minimizing the ${\sf CVaR}$ of the loss $-R$ at a confidence level $\gamma$, i.e., ${\sf CVaR}_\gamma[-W_0R]$, that is altogether controlled by a parametric policy with a decision vector $\Theta$. Thus, Problem \eqref{eq:utility} becomes:
\begin{equation}
    \begin{aligned}
    &\min_\Theta\quad {\sf CVaR}_\gamma[-W_0R(\Theta)]
    \end{aligned}    
    \label{eq:opt}
\end{equation}
By the discussion in \cite{sarykalin2008value}, we state the following Lemma.
\begin{lemma}
If $R(\Theta)$ is concave, then, Problem \eqref{eq:opt} is convex. \label{lemma:R}
\end{lemma}
\par In this work, we generally consider a single-asset trader under the assumption of negligible impact of the trader's positions on the asset's price, i.e., trading volume is significantly less than that of the overall market. Also, it is worth noting that for the special case when $\gamma=0$, the agent is risk-neutral, and Problem \eqref{eq:opt}  becomes equivalent to 
\begin{equation}
    \begin{aligned}
    &\max_\Theta\quad \mathbb{E}[W_0 R(\Theta)]
    \end{aligned}
\end{equation}
\begin{figure}[!t]
     \centering
     \includegraphics[width=0.9\linewidth,bb=0 0 420 230]{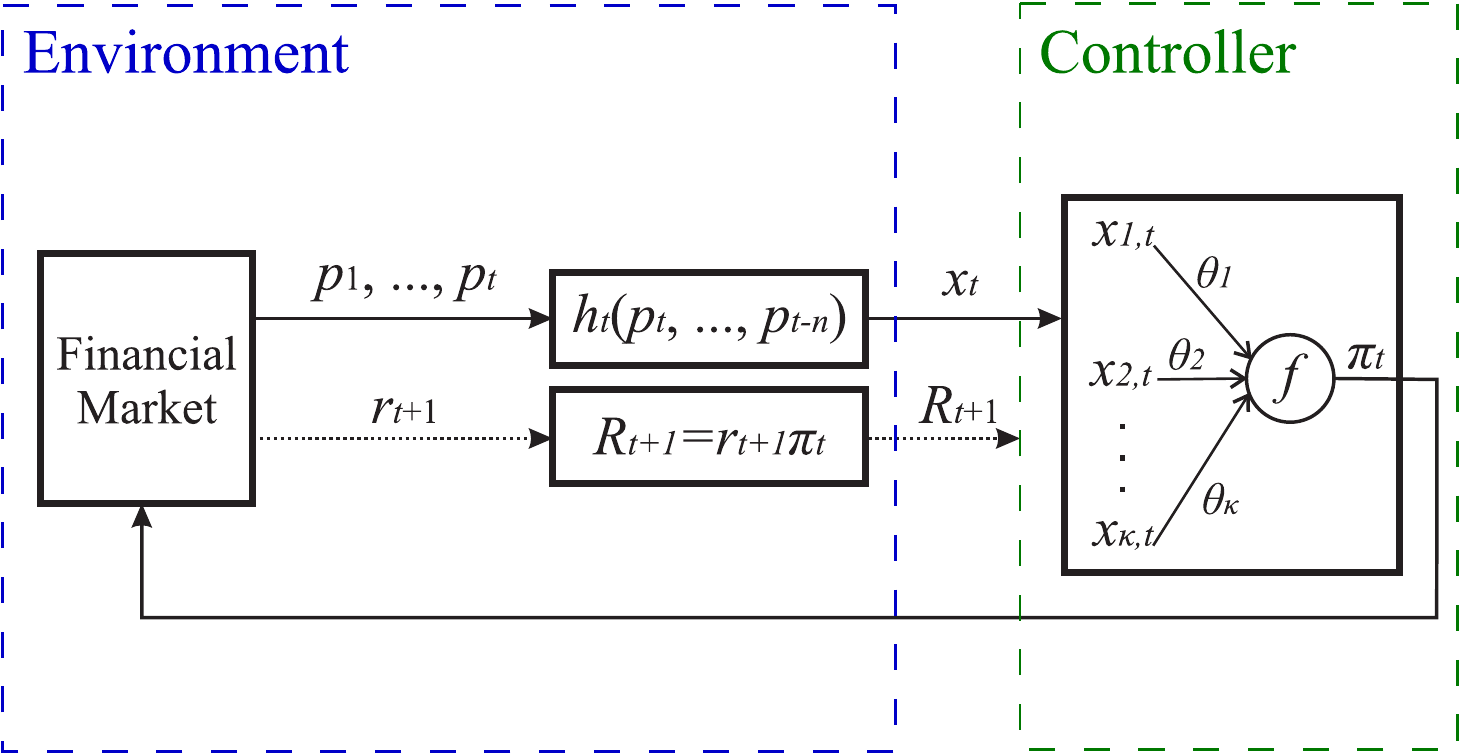}
     \caption{Trading system's block diagram. Solid signals represent sensing the environment and taking action during $[t,t+1)$. Dashed signals represent receiving a reward at $t+1$}\label{fig:systemDiagram}
\end{figure}
Our goal is to search for a decision vector $\Theta$ that solves Problem \eqref{eq:opt}. Using {\sf CVaR}, we have a risk-sensitive representation of the generated profits and incurred losses during the investment horizon $T$. 
\par In Fig. \ref{fig:systemDiagram}, we visualize the sequence of decisions and flow of information. At each time step $t$, the controller senses the environment through $x_t$, generates a trading signal following the deterministic policy $\pi_t$, and receives a reinforcing signal $R_{t+1}$ at the immediate next time step. Hence, the reward signal at each step can be described by:
\begin{equation}
    \label{eq:reward}
    R_{t+1} = \pi_{t}r_{t+1}
\end{equation}
Where $r_{t+1}=p_{t+1}/p_t-1$ is the normalized price return of an asset. For the policy $\pi_t$ that is followed by our agent, we model it by a single-unit neural network that is defined by:
\begin{equation}
    \label{eq:policy}
    \pi_t = f[g_t(x_t;\Theta)]
\end{equation}
Where:
\begin{equation}
    g_t(x_t;\Theta)=\Theta^\textrm{T} x_t
\end{equation}
$f(\cdot)$ is an activation function of a neural network's neuron, $x_t=[x_{1,t}\quad x_{2,t}\quad \cdots\quad x_{\kappa,t}\quad 1]^\textrm{T} \in \mathbb{R}^{\kappa+1}$ represents the system with $\kappa$ states at time $t$, and $\Theta=[\theta_1\quad\theta_2\quad\cdots\quad\theta_\kappa\quad \theta_{\kappa+1}] \in \mathbb{R}^{\kappa+1}$ is the decision vector where $\theta_{\kappa+1}$ is the bias parameter. From the reward \eqref{eq:reward} and the policy in \eqref{eq:policy}, we can describe $R_{t+1}$ explicitly as a function of $\Theta$ and the asset's return as:
\begin{align}
    \label{eq:reward_theta}
    R_{t+1}(\Theta) = f[g_{t}(x_{t};\Theta)]r_{t+1}
\end{align}
We can see that, under the general case of $\kappa$ states, the reward in \eqref{eq:reward_theta} is a multivariate function such that $R_{t+1}(\Theta):\mathbb{R}^{\kappa+1}\rightarrow\mathbb{R}$. Note that the definition of the reward in \eqref{eq:reward_theta} forms the basis of a "batch" algorithm where the decision variable is only updated after a full sweep over the data within the investment horizon $T$. Similar to Moody and Saffell in \cite{moody2001learning}, we do online parameter updates, for consistency with real finance applications, and we have an adaptive trading system that can robustly detect financial market regime-switching. Such practical considerations will be discussed later in Section \ref{sec:experiment}.
The following result, whose proof can be found in the Appendix, provides sufficient conditions for the convexity of Problem \eqref{eq:opt}. 
\begin{theorem}
Problem \eqref{eq:opt} is convex in $\Theta$ if, for each $t$, one of the following conditions hold: 
\begin{enumerate}
    \item $r_{t+1} \geq 0$ and $f$ is concave.
    \item $r_{t+1} \leq 0$ and $f$ is convex.
\end{enumerate}
    \label{convex}
\end{theorem}
Theorem \ref{convex} elucidates that the convexity of the problem depends on the asset price returns and the convexity of the selected activation. In particular, to ensure that one of the conditions hold at each $t$, one must have an activation that can be either concave or convex, depending on the sign of $r_{t+1}$. The hyperbolic tangent (or sigmoid function in long-only trading) and linear activations are obvious selections that may achieve those requirements and are suitable for trading. Satisfying one of the conditions in Theorem \ref{convex} is certain with the linear activation. However, with the hyperbolic tangent, one can fulfill one of the conditions only under the case of profitable trades, i.e., $\pi_t,r_{t+1}\geq0$ or $\pi_t,r_{t+1}\leq0$, otherwise the problem becomes non-convex. To enforce convexity, one may consider clipping the reward function, i.e., $R_{t+1}=0$, when the trade is unprofitable. With that, however, the controller only receives feedback about its sound decisions. Nevertheless, we will show later in Section \ref{sec:experiment} that both selections are effective under our formulation. We emphasize here that, as we mentioned in Section \ref{sec:relWork}, convexity is rare in quantitative trading, and it has not been explored before under the DRL notion.
\par Noteworthy, so far, our formulation of the reward function considers the ideal case in which all associated trading frictions, e.g., transaction costs, are neglected. Incorporating such factors into the model requires perturbations of the reward function. However, as we will show later, such perturbations do not violate convexity.
\subsection{Online Computation of the Optimal Policy}

\par The market environment can be represented by macroeconomic variables \cite{humpe2009can} or technical indicators \cite{neely2014forecasting}. To simplify the discussion, one can consider representing it by a series of past filtered price returns, i.e., $x_t=[y_t, y_{t-1},\dots y_{t-n},1]^{\textrm{T}}$ where $y_t$ is the return of the filtered price signal at time $t$. We note here that the use of past returns in quantitative finance applications is commonplace; see for example \cite{black1992global, demiguel2009generalized}. 
For the policy activation $f(\cdot)$, we use a bounded linear function to represent the allowable range of the trading signal, i.e., $[-1,1]$. As we discussed, convexity is guaranteed with linear activation while it is not with the hyperbolic tangent and sigmoid functions, which might make the system very sensitive to initialization. Furthermore, as we demonstrate later, this allows us to consider trading frictions while preserving the convexity (see Section \ref{marketFriction}). We use a unity investment capital $W_0=1$. Therefore, the received reward signal by the agent is purely the normalized return of the traded asset scaled by the trading signal $\pi_t$, with the final wealth, or cumulative return over the investment horizon is $W_T=W_0R$. 
\par As we discussed earlier, we aim to learn the optimal policy in an online manner where the decision $\pi_t$ is immediately evaluated when receiving $R_{t+1}$. 
We now show how to compute the objective in Problem \eqref{eq:opt} at each time step and how we recursively learn the optimal policy. We use the subscript $i$ to represent our iterative computation and distinguish it from the market time step $t$.
\par Our objective described in Problem \eqref{eq:opt} mandates computing ${\sf CVaR}_\gamma$ to solve it. Since the reward samples arrive sequentially during the trading process, we can estimate ${\sf CVaR}_\gamma$ at each time step using all captured rewards up to $t+1$, i.e., $R_1, R_2, ..., R_{t+1}$ and we denote it by ${\sf CVaR}_{\gamma,t+1}$. Our estimation follows the Monte Carlo approach proposed by Hong and Liu in \cite{hong2011monte}, which proposed the following convergent Monte Carlo estimation of ${\sf CVaR}_{\gamma,t+1}[-W_0R(\Theta_t)]$, denoted by $\hat{c}_{\gamma,i}$ (We drop the constant $W_0$ for brevity):
\begin{equation}
    \mathrm{\hat{c}}_{\gamma,i} = \mathrm{\hat{v}}_{\gamma,i}+\frac{1}{(t+1)(1-\gamma)}\sum_{j=1}^{t+1}[-R_j-\mathrm{\hat{v}}_{\gamma,i}]^+
    \label{eq:estCVaR}
\end{equation}
Where $[x]^+ = \max(x,0)$ and $\mathrm{\hat{v}}_{\gamma,i}$ is the estimation of ${\sf VaR}_{\gamma,t+1}[-W_0R(\Theta_t)]$, which is given by:
\begin{equation}
    \mathrm{\hat{v}}_{\gamma,i} = -R_{\lceil(t+1)\gamma\rceil:t+1}
    \label{eq:estVaR}
\end{equation}
Where $-R_{\lceil k:(t+1)\rceil}$ is the $k^\mathrm{th}$ order statistic from the $(t+1)$ samples. Once the estimation in \eqref{eq:estCVaR} is computed, we can directly apply stochastic gradient descent (SGD) to update the policy's parameters $\Theta$:
\begin{equation}
    \Theta_{i} = \Theta_{i-1} - \alpha\nabla_{\Theta_{i-1}}\mathrm{\hat{c}}_{\gamma,i}
    \label{eq:param_update}
\end{equation}
Where $\alpha\in(0,1)$ is the learning rate. We need to point out that optimization with SGD has an attractive privilege in our problem. Since SGD provides “noisy” gradient information, it enhances the exploratory behavior of the controller. That behavior can be altered by tuning the learning rate. However, the online update of $\Theta$ governed by \eqref{eq:estCVaR}, \eqref{eq:estVaR}, and \eqref{eq:param_update} has two challenges. The first challenge is vanishing gradient information in the case of $-R_{t+1}<\mathrm{\hat{v}}_{\gamma,i}$ where $\mathrm{\hat{c}}_{\gamma,i}$ is no longer a function of $\Theta_{i-1}$. That problem becomes even worse as $\gamma\rightarrow1$ since the number of samples that fall below $\mathrm{\hat{v}}_{\gamma,i}$ grows with higher $\gamma$. To overcome the first challenge, we add the second norm of $\Theta$ as a regularization term to Problem \eqref{eq:opt}; thus the problem becomes:
\begin{equation}
    \begin{aligned}
    &\min_\Theta\quad {\sf CVaR}_\gamma[-W_0R(\Theta)] + \lambda\|\Theta\|_2
    \end{aligned}    
    \label{eq:opt_reg}
\end{equation}
Where $\lambda\in[0,1]$ is the regularization constant. The regularization term guarantees the availability of gradient information in all time steps. Moreover, as stated by Bishop in \cite{bishop1995training}, the regularization term enhances the generalization performance of learning systems with objectives that do not involve error minimization. The second challenge we have is a biased estimate  $\mathrm{\hat{c}}_{\gamma,i}$ as $t$ increases. Having a biased estimate is problematic in our particular application since we expect the distribution of the reward to vary over time due to the non-stationary behavior of the financial market. To circumvent the second issue, we propose an adaptive estimation for ${\sf CVaR}_{\gamma,t+1}$. In that sense, the estimate at each step is derived using only a predetermined $N+2$ past samples of the reward. Thus, the estimates in \eqref{eq:estCVaR} and \eqref{eq:estVaR} become:
\begin{align}
    \mathrm{\hat{c}}_{\gamma,i} &= \mathrm{\hat{v}}_{\gamma,i}+\frac{1}{(N+2)(1-\gamma)}\sum_{j=t-N}^{t+1}[-R_j-\mathrm{\hat{v}}_{\gamma,i}]^+    \label{eq:estCVaR_N}\\    \mathrm{\hat{v}}_{\gamma,i} &= -\bar{R}_{\lceil (N+2)\gamma\rceil:N+2} \label{eq:estVaR_N}
\end{align}
Where $\bar{R} = [R_{t-N}\quad R_{t-N-1}\quad \cdots \quad R_{t+1}]$. With that approach, we ensure the estimates are adaptive and reflect any potential market-regime switching. In addition, the proposed approach enhances the computational efficiency of deriving the estimates since with the original formulation the computational cost increases with $t$. We now can update $\Theta$ recursively as follows:
\begin{equation}
    \Theta_{i} = \Theta_{i-1} - \alpha\nabla_{\Theta_{i-1}}\big(\mathrm{\hat{c}}_{\gamma,i}+ \lambda\|\Theta_{i-1}\|_2\big)
    \label{eq:param_update_reg}
\end{equation}
The estimates in \eqref{eq:estCVaR_N}, \eqref{eq:estVaR_N} and the update rule in \eqref{eq:param_update_reg} form the basis of our online learning algorithm that optimizes Problem \eqref{eq:opt} under frictionless setup. 
\begin{algorithm}[!t]
\caption{$\sf CVaR$-Sensitive Online Optimal Policy}
\begin{algorithmic}
\STATE \textbf{Select:} $W_0,\gamma,\alpha,\lambda,N,n,I$
\STATE \textbf{Initialize:} $\Theta_0,\vartheta_0,\eta_0$ and $t=0, i=1$
\FOR{$t<T$}
\STATE \textbf{Environment}
\STATE \textit{Sense the environment}: $x_t \leftarrow h_t(p_t,...,p_{t-n})$
\STATE \textit{Take action}: $\pi_t\leftarrow f(\Theta^\textrm{T}_t x_t)$
\STATE \textit{Observe reward}: $R_{t+1} \leftarrow \pi_tr_{t+1}-\delta\vartheta_t$
\STATE \textbf{Computation}
\quad\FOR{$i\leq I$}
\STATE \textit{Compute ${\sf VaR}_{\gamma,t+1}$}: $\mathrm{\hat{v}}_{\gamma,i} \leftarrow$ \eqref{eq:estVaR_N}
\STATE \textit{Compute ${\sf CVaR}_{\gamma,t+1}$}: $\mathrm{\hat{c}}_{\gamma,i} \leftarrow $ \eqref{eq:estCVaR_N}
\STATE \textit{Update decision vector}: $\Theta_{i} \leftarrow$ \eqref{eq:param_update_reg_const}
\STATE \textit{Update aux. decision variable}: $\vartheta_{i} \leftarrow$ \eqref{eq:vartheta_update_reg_const}
\STATE $\eta_i\leftarrow\eta_{i-1}/{i}$
\ENDFOR
\STATE $\Theta_{t+1} \leftarrow \Theta_i$
\STATE $\vartheta_{t+1} \leftarrow \vartheta_i$
\ENDFOR
\end{algorithmic}
\label{algo:onlinePolicyAlgo}
\end{algorithm}
Due to the convexity of our problem, our experiments demonstrate that the learning algorithm can adaptively discover a lucrative strategy through a single run over the data. Besides that, the proposed model is significantly more efficient than other methods in computing the solution. To the best of our knowledge, this is the first work that proposes a trading system under a machine learning paradigm that learns on the go with well-performance. 

\subsection{Including Market Frictions} \label{marketFriction}
In the finance literature, trading frictions, such as transaction costs, can be incorporated into the problem in several ways. For example, Davis and Norman in \cite{davis1990portfolio} modeled the transaction charges as a linear function of the transacted assets amount in a portfolio selection problem. Morton and Pliska in \cite{morton1995optimal}, on the other hand, implemented a fixed cost that depends on the portfolio value. Here we follow the approach in \cite{davis1990portfolio} since it is consistent with the pioneering DRL trading systems \cite{moody2001learning,deng2016deep}. To this end, the reward function in \eqref{eq:reward} can be altered as follows:
\begin{equation}
     R_{t+1}(\Theta_t) = \pi_{t}r_{t+1} - \delta |\pi_{t} - \pi_{t-1}|
     \label{eq:reward_tc}
\end{equation}
Where $\delta$ is a fixed percentage that counts for half of the round-trip cost (only opening or closing a position), and the argument inside the absolute value represents the traded amount of the asset at time $t$. With linear activation function, the argument inside the absolute value is linear in $\Theta_t$. Thus, with the use of an auxiliary decision variable $\vartheta_t$, we can reformulate the absolute value and add linear constraints to Problem \eqref{eq:opt_reg}. Hence, the reward function becomes:
\begin{equation}
    R_{t+1}(\Theta_t,\vartheta_t) = \pi_{t}r_{t+1} - \delta \vartheta_t
    \label{eq:auxReward}
\end{equation}
While Problem \eqref{eq:opt_reg} is now reformulated as:
\begin{equation}
    \begin{aligned}
    &\min_{\Theta,\vartheta}\quad {\sf CVaR}_\gamma[-W_0R(\Theta,\vartheta)] + \lambda\|\Theta\|_2\\
    &\quad\textrm{s.t.}\quad \pi_{t} - \pi_{t-1}- \vartheta\leq 0\\
    &\quad\quad -\pi_{t} + \pi_{t-1}- \vartheta\leq 0
    \end{aligned}    
    \label{eq:opt_reg_const}
\end{equation}
Since Problem \eqref{eq:opt_reg_const} is convex, we unfold it using the well-known log barrier method \cite{boyd2004convex}. We thus can have an augmented objective described by:
\begin{flalign}
    \begin{aligned}
    &\min_{\Theta,\vartheta}\quad {\sf CVaR}_\gamma[-W_0R(\Theta,\vartheta)]+  \lambda\|\Theta\|_2\\
    &\qquad\quad  +\frac{1}{\eta}\Big[\log\big(\pi_{t} - \pi_{t-1} + \vartheta\big)\\
    &\qquad\qquad  +\log\big(-\pi_{t} + \pi_{t-1}+\vartheta\big)\Big]
    \end{aligned}
\end{flalign}    
\label{eq:opt_aug}
Where $\eta>0$ is a parameter that determines the accuracy of the approximation to the original Problem \eqref{eq:opt_reg_const}, and it should grow with time during computing the solution, i.e., $1/\eta\rightarrow0$ as $i\rightarrow\infty$. 
We now can compute $\Theta$ update recursively as follows:
\begin{equation}
    \begin{aligned}
    &\Theta_{i} = \Theta_{i-1} -\\ &\alpha\nabla_{\Theta_{i-1}}\bigg\{\mathrm{\hat{c}}_{\gamma,i}+ \lambda\|\Theta_{i-1}\|_2+\frac{1}{\eta_i}\Big[\log\big(\pi_{i-1} - \pi_{i-2} + \vartheta_{i-1}\big)\\
    &+\log\big(-\pi_{i-1} + \pi_{i-2}+\vartheta_{i-1}\big)\Big]\bigg\}
    \label{eq:param_update_reg_const}
    \end{aligned}
\end{equation}
We similarly update the auxiliary variable $\vartheta$:
\begin{equation}
    \begin{aligned}
    &\vartheta_{i} = \vartheta_{i-1} -\\ &\alpha\frac{\partial}{\partial\vartheta_{i-1}}\bigg\{\mathrm{\hat{c}}_{\gamma,i}+ \lambda\|\Theta_{i-1}\|_2+\frac{1}{\eta_i}\Big[\log\big(\pi_{i-1} - \pi_{i-2} + \vartheta_{i-1}\big)\\
    &+\log\big(-\pi_{i-1} + \pi_{i-2}+\vartheta_{i-1}\big)\Big]\bigg\}
    \label{eq:vartheta_update_reg_const}
    \end{aligned}
\end{equation}
The estimates in \eqref{eq:estCVaR_N} and \eqref{eq:estVaR_N} considering the reward in \eqref{eq:auxReward} with the update rules in  \eqref{eq:param_update_reg_const} and \eqref{eq:vartheta_update_reg_const} form the basis of our online algorithm under frictional market, which is depicted in Algorithm \ref{algo:onlinePolicyAlgo}.

\begin{figure}[!t]
     \centering
     \includegraphics[width=1\linewidth,bb=0 0 450 275]{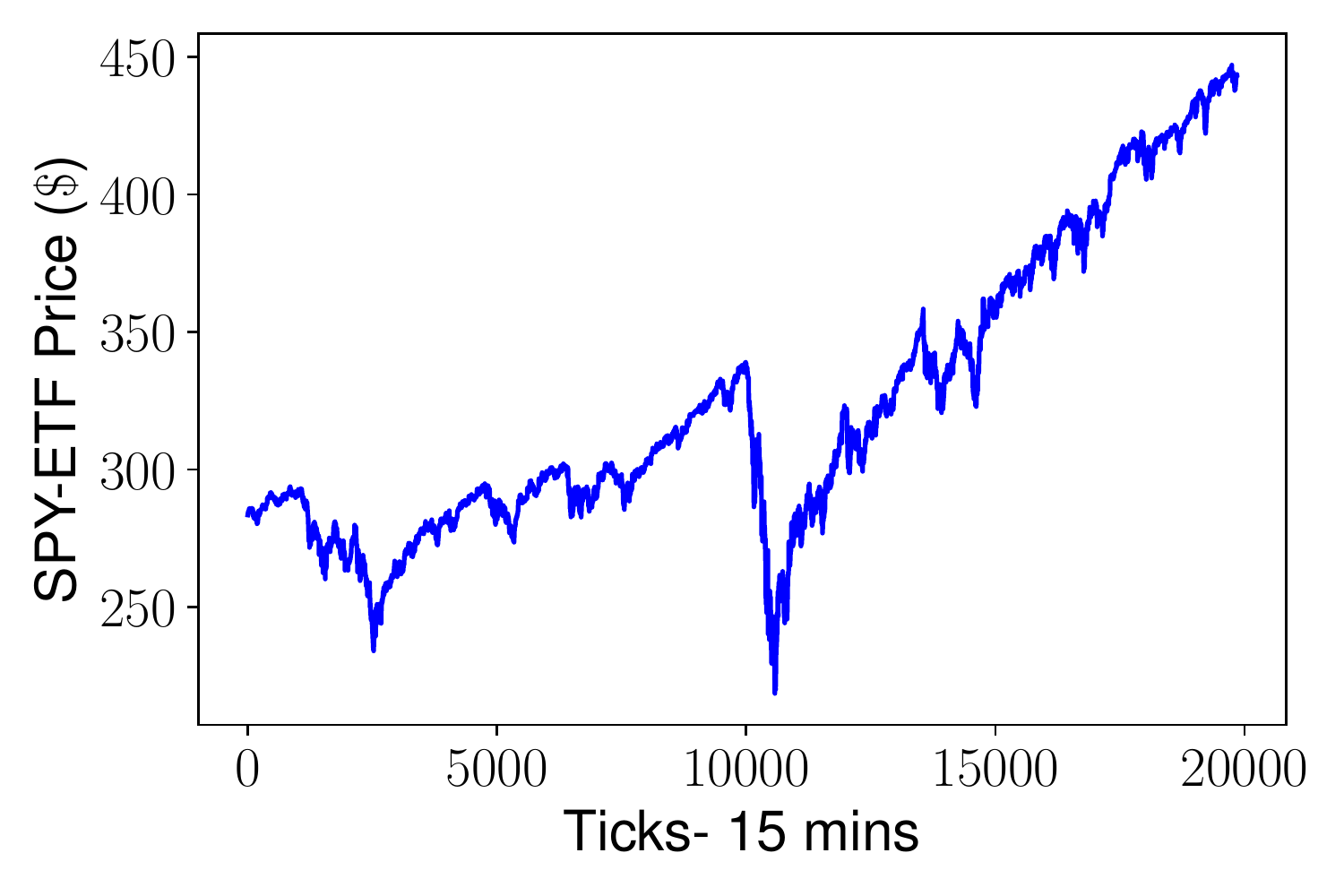}
     \caption{US SPDR trust fund price signal from 20/08/2018 to 06/08/2021}\label{fig:SPY}
\end{figure}
\begin{figure*}[!tb]
    \centering
    \subfigure[Overall returns Vs. $\gamma$]
    {
        \includegraphics[width=.3\linewidth,bb= 0 0 365 365]{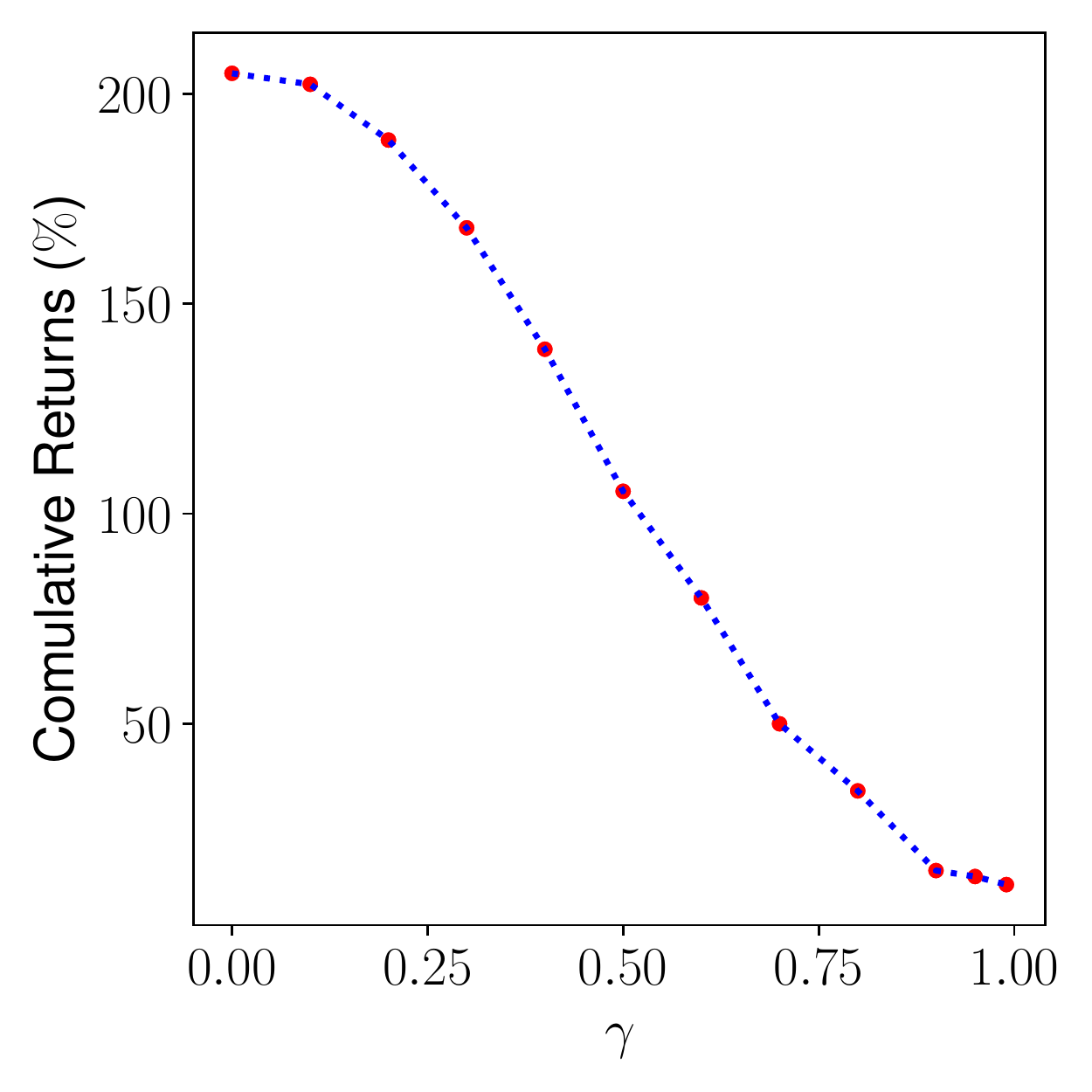}
        \label{fig:returnsGamma}
    }
    \subfigure[Negative returns variance Vs. $\gamma$]
    {
        \includegraphics[width=.3\linewidth,bb=0 0 365 365]{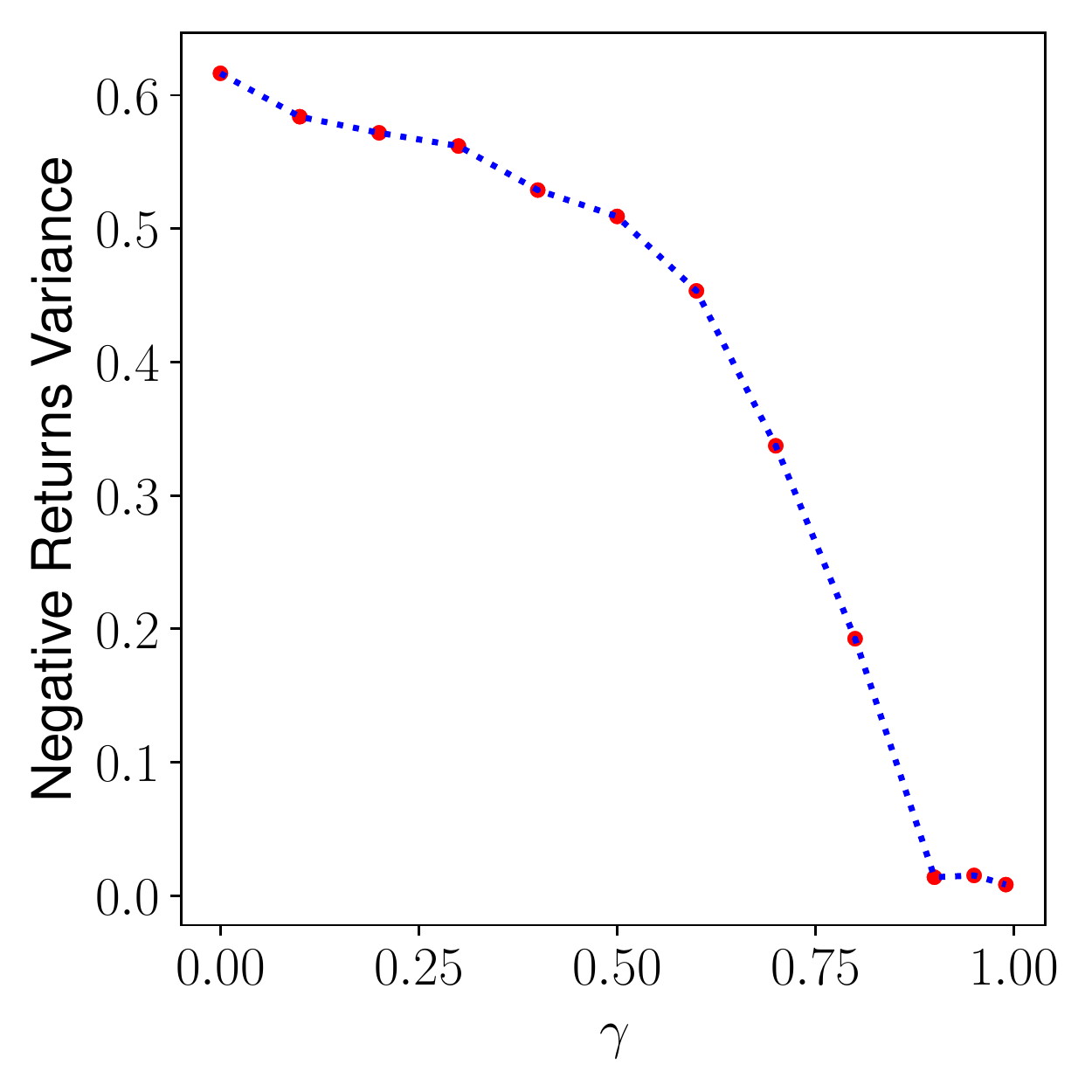}
        \label{fig:negRetGamma}
    }
    \subfigure[Max. drawdown Vs. $\gamma$]
    {
        \includegraphics[width=.3\linewidth,bb= 0 0 365 364]{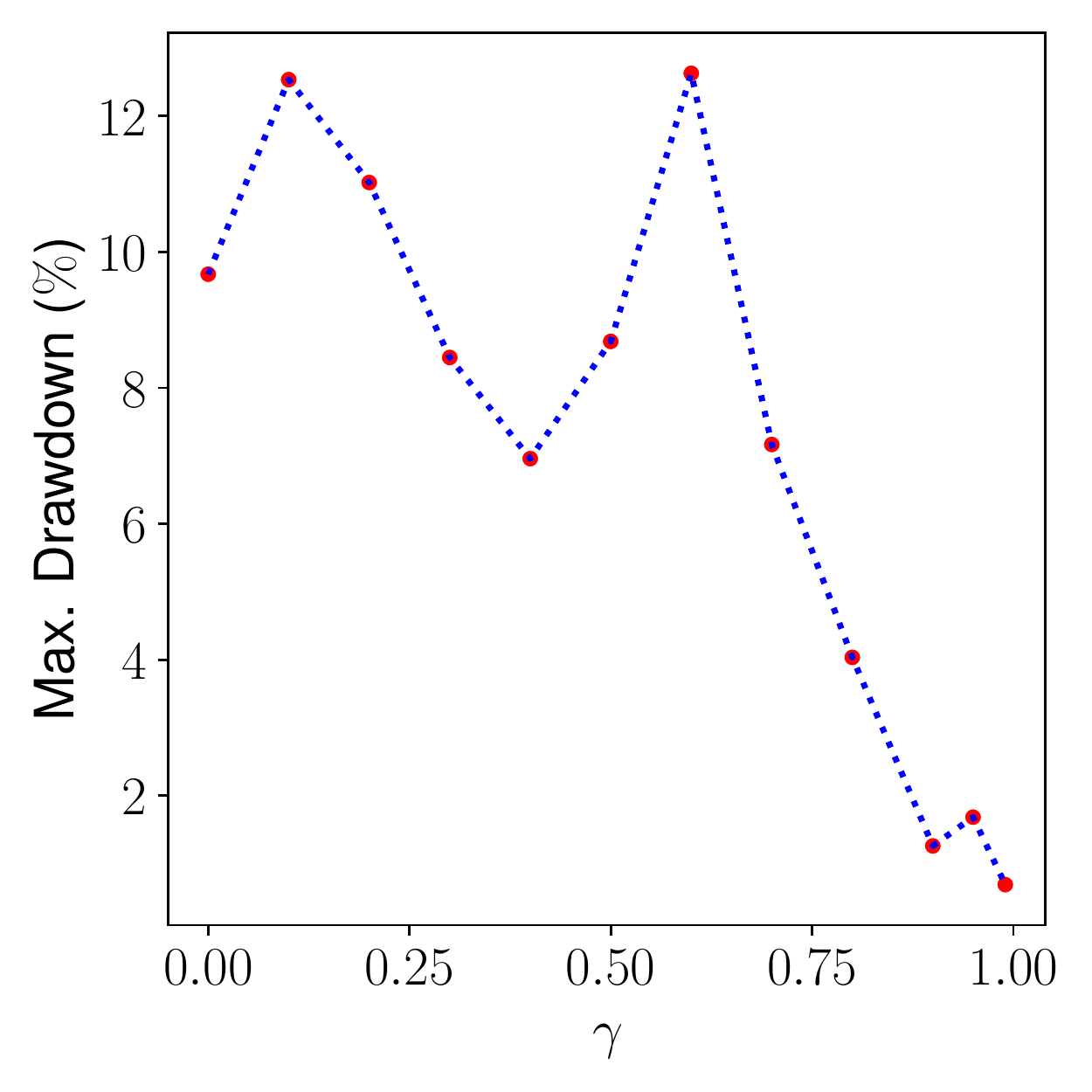}
        \label{fig:mddGamma}
    }
    \caption{Results of frictionless market with different risk-aversion parameter ($\gamma$)}
    \label{fig:gammaVer}
\end{figure*}

\begin{figure*}[!tb]
    \centering
    \subfigure[${\sf CVaR}_{\gamma=0}$ $-$ Risk-neutral]
    {
        \includegraphics[width=.3\linewidth,bb= 0 0 365 365]{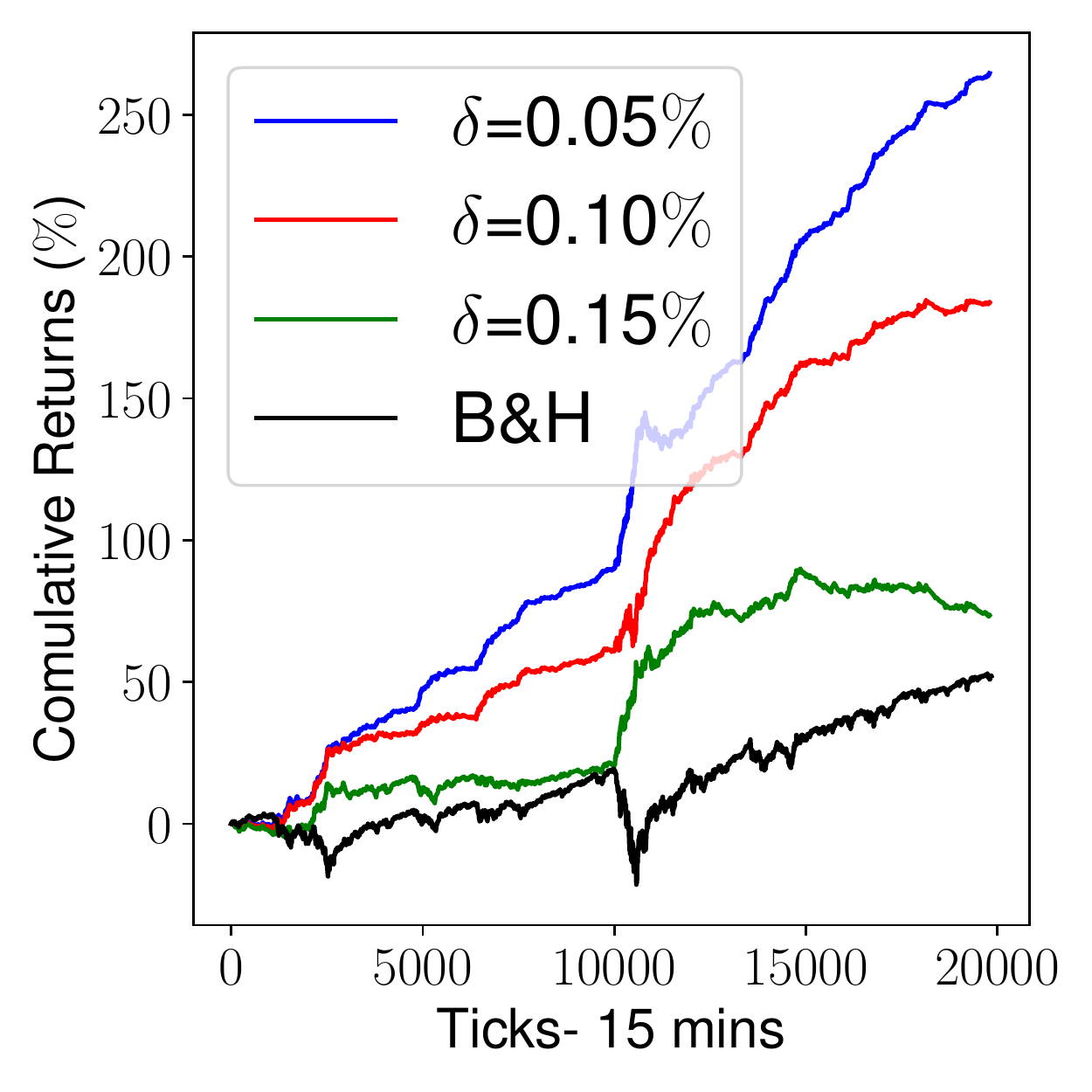}
        \label{fig:cumRetExp}
    }
    \subfigure[${\sf CVaR}_{\gamma=0.9}$ $-$ Risk-averse]
    {
        \includegraphics[width=.3\linewidth,bb = 0 0 365 365]{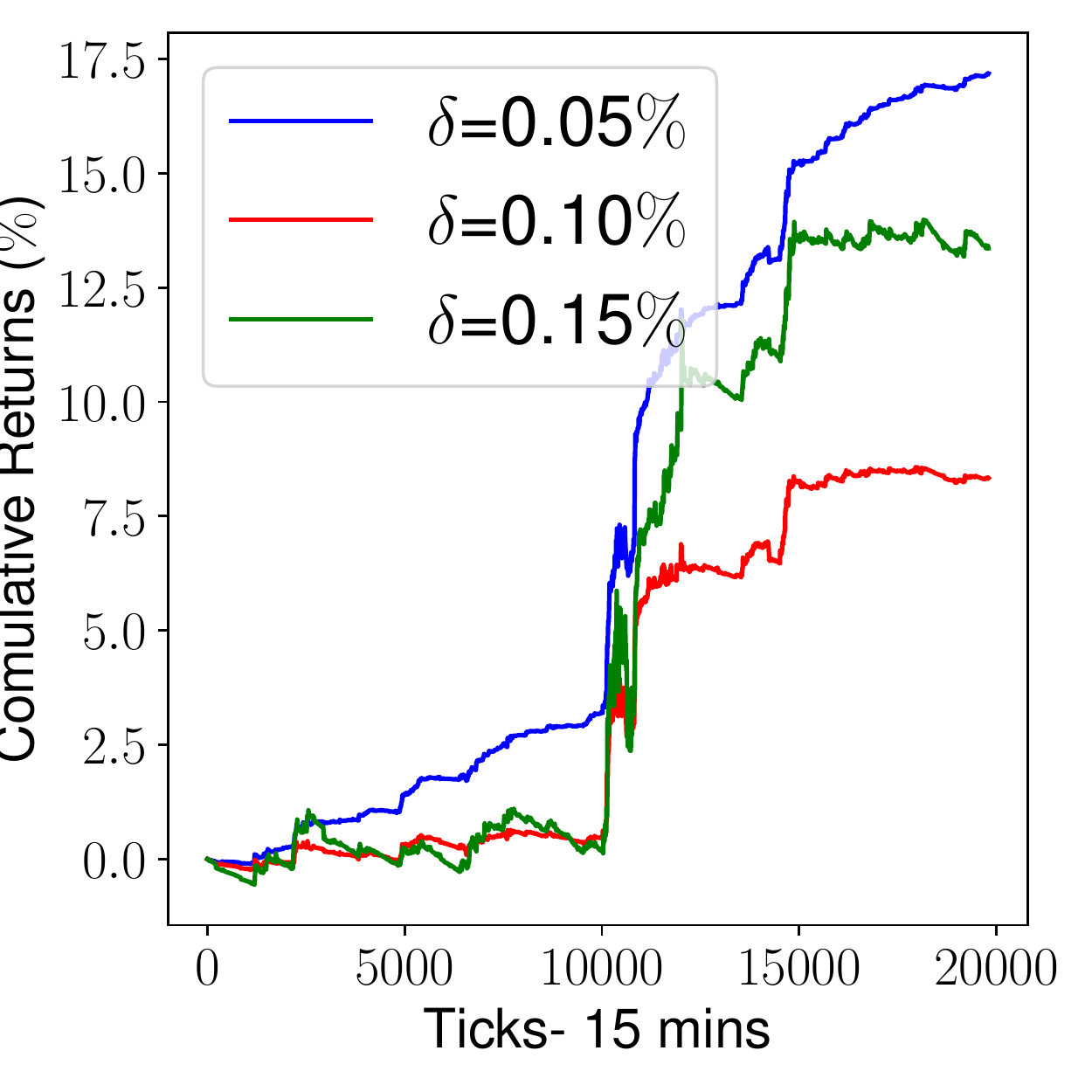}
        \label{fig:cumRetCVaR}
    }
     \subfigure[Linear Vs. nonlinear activation. Linear activation ensures convexity (see Theorem \ref{convex})]
    {
        \includegraphics[width=.3\linewidth,bb= 0 0 365 365]{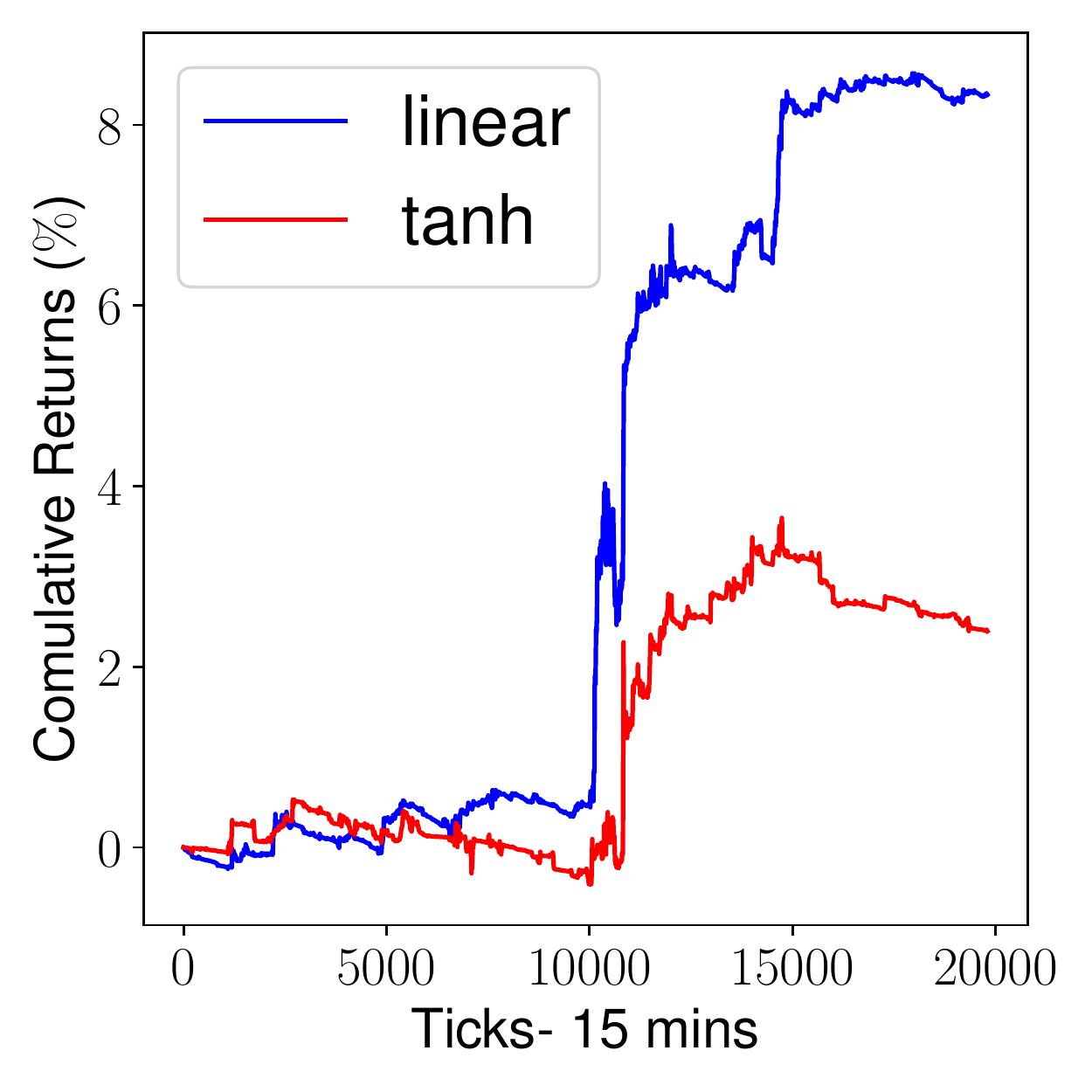}
        \label{fig:tanhLinear}
    }
    \caption{Cumulative returns under various setups}
    \label{fig:cumRetExp-CVaR}
\end{figure*}

\section{Experimental Results}\label{sec:experiment}
\par We assess our proposed model on actual financial market data. We, in particular, use one of the largest US trust funds, the SPDR (ticker: SPY) \cite{poterba2002exchange} price signal from 06/08/2018 to 20/08/2021 with a trading interval of $\Delta t = 15$ minutes. That forms an investment horizon of $T = 20000$ steps. As indicated in the SPY price signal shown in Fig. \ref{fig:SPY}, the price experienced severe fluctuation and had several market regime-switching during that period. Therefore, we believe that the selected out-of-sample data is reasonable for testing the robustness of our proposed trading system. Our numerical experiments were conducted using a standard computer equipped with a 3.2-GHz 8-core processor and 16-GB RAM, which makes our approach attractive and easily implementable in practice. 
\par We show the results of our trading system over the out-of-sample data under a different selection of the risk-aversion parameter $\gamma$. We also assess the trading system under conservative transaction costs, i.e., $\delta=0.05,0.1$ and $0.15\%$ (equivalent to a round-trip charges of $0.1, 0.2$ and $0.3\%$), to account for other frictions associated with trading, including spread and slippage costs. 
\par First, in Fig. \ref{fig:gammaVer}, we illustrate the algorithm's effectiveness in risk management and validate that it could uncover lucrative trading strategies under a frictionless market. Over the investment horizon, Fig. \ref{fig:returnsGamma} shows the overall returns, Fig. \ref{fig:negRetGamma} illustrates the variance of negative returns, and Fig. \ref{fig:mddGamma} demonstrates the MDD. Those figures show the results at different risk preferences $\gamma$. Note that we report the variance of negative returns rather than the overall since ${\sf CVaR}_\gamma$ is concerned about large incurred losses that fall within the $(1-\gamma)$-percentile of the reward distribution. Consequently, we can see in Fig. \ref{fig:negRetGamma} a consistent pattern of variance reduction as $\gamma\rightarrow1$. Further, we see in Fig. \ref{fig:returnsGamma} that the highest total return is achieved when the objective is the reward expectation (i.e., ${\sf CVaR}_{\gamma=0}$).  However, this high return comes at the cost of large MDD, around $10\%$. Although we see the lowest return at $\gamma=0.99$, it has the lowest MDD at about $0.7\%$. Moreover, for $\gamma\geq0.9$, both the variance and returns are almost steady at around $0.015$ and $14\%$, respectively. That can be attributed to the number of samples used to estimate ${\sf CVaR}_\gamma$ where we used 50 samples for $\gamma=0.9, 0.95$ and 100 samples for $\gamma=0.99$. Indeed, the result of frictionless trading proves the algorithm's subtlety in extracting trading schemes that strictly comply with the investor's risk preference. 
\par To gain deeper insights into proposed model's practicality, we exhibit the cumulative returns over the investment horizon under different transaction costs in Fig. \ref{fig:cumRetExp-CVaR} where Fig. \ref{fig:cumRetExp} shows the results of expectation as an objective, and Fig. \ref{fig:cumRetCVaR} shows the case of risk-adjusted returns with ${\sf CVaR}_{\gamma=0.9}$. Under the expectation case illustrated in Fig. \ref{fig:cumRetExp}, we observe that the trading system was learning in the first 1000 steps where no profits were made before that point. Note that with higher round-trip ($2\delta$) trading charges of 0.3\%, the system even took a longer time than the case of 0.1 and 0.2\% to learn and did not discover a profitable trading scheme before the $2000^{th}$ step. Most importantly, we can notice the robustness of the controller within its achieved profits through detecting the primary market-regime switching occurred at the significant price pullback (around step 10,000 in Fig. \ref{fig:SPY}) and price rise (step 10,500 in Fig. \ref{fig:SPY}). Further, our trading system outperforms the baseline strategy of buying and holding (B\&H) the asset over the investment horizon under all considered trading chargers. For the case of risk-adjusted returns with $\gamma=0.9$ shown in Fig. \ref{fig:cumRetCVaR}, the controller seems conservative in trading unless it uncovers a clear pattern within the price signal, and this justifies why its total return is lower than the case of expectation. In addition, with $\gamma=0.9$, the controller tries to lower its trading positions to avoid significant losses, and this indeed comes at the cost of lower total returns over the investment horizon. To this end, we see that the proposed algorithm proves its practicality where it shows that it can have lucrative strategies even in the case of most conservative trading costs. At the same time, it avoids risky decisions at higher values of the risk-aversion parameter.
\par Finally, in Fig. \ref{fig:tanhLinear}, we compare the performance of a policy with linear activation where the problem is convex against the case of non-convex formulation with the hyperbolic tangent as an activation. Under that case, we consider optimizing ${\sf CVaR}_{\gamma=0.9}$ at a round-trip transaction cost of $0.2\%$, with $\delta=0.1$. We can see the privilege of convex formulation, where the total return of the linear policy is about fourfold of the hyperbolic tangent.
\section{Generalizations} \label{general}
Our proposed convex Problem \eqref{eq:opt} can easily be generalized to consider risk-constrained and assets allocation problems. Here we discuss the theoretical setup of those generalizations, but we leave validating them through numerical experiments for future work. In some instances, a trader may seek to maximize the expectation of the returns amid avoidance of potentially large losses. In that sense, we can consider the following alternative to Problem \eqref{eq:opt}:
\begin{equation}
    \begin{aligned}
    &\max_\Theta\quad \mathbb{E}[W_0R(\Theta)]\\
    &\quad\mathrm{ s.t.}\quad \mathrm{{\sf CVaR}}_\gamma[-W_0R(\Theta)] \leq \mu_\beta
    \end{aligned}    
    \label{eq:opt_general1}
\end{equation}
Where $\mu_\beta$ represents the investor's tolerable expected loss that may exceed $\mathrm{VaR}_\gamma$. 
The conditions for the convexity of Problem \eqref{eq:opt_general1} are the same as those in Problem \eqref{eq:opt} and are given by Theorem \ref{convex}. Noteworthy, recent developments in the literature of risk-sensitive constrained optimization makes solving Problem \eqref{eq:opt_general1} achievable even in more general setups, see for example \cite{madavan2021stochastic}. 
\par The second, and more important generalization, is extending the formulation to an asset allocation problem. In that setup, we consider a long-only policy where short selling is not allowed, i.e., $\pi_t\in[0,1]$. Also, the controller follows a policy that can re-balance its investment capital at each time step among risky (e.g., stocks) or risk-less (e.g., Treasury yields) assets. We then can reformulate the reward function as follows:
\begin{equation}
    \label{eq:reward_portfolio}
    R_{t+1}^p = \pi_{t}r_{t+1} + (1-\pi_{t})r^f_{t+1}
\end{equation}
Where $r_{t+1}$ and $r_{t+1}^f$ represent the risky and risk-less asset returns, respectively. We then optimize Problem \eqref{eq:opt} based on the reward function in \eqref{eq:reward_portfolio}. Similar to Theorem \ref{convex}, we have the following result, whose proof can be found in the Appendix.
\begin{theorem}
With the reward function in \eqref{eq:reward_portfolio} in a long-only setup, Problem \eqref{eq:opt} is convex in $\Theta_t$ if, for each $t$, one of the following conditions is true: 
\begin{enumerate}
    \item $r_{t+1}-r_{t+1}^f \geq 0$ and $f$ is concave.
    \item $r_{t+1}-r_{t+1}^f \leq 0$ and $f$ is convex.
\end{enumerate}
    \label{convex_general2}
\end{theorem}
\section{Conclusion}
In this manuscript, we demonstrated the effectiveness of convex trading systems with $\mathrm{\sf CVaR}$ as a risk-adjusted performance measure following the concept of DRL. Due to convexity, the proposed controller illustrated its capability in learning and updating its policy in a pure online manner. In that sense, the proposed approach showed its robustness in detecting primary market-regime switching while generating profits under a realistic frictional market for three years. Further, the numerical experiment revealed that the algorithm effectively manages risk based on the investor's risk preference $\gamma$. The trading system could outperform the market under the risk-neutral case, i.e., $\gamma=0$, while it could achieve a minimal MDD with the most risk-averse trader of $\gamma=0.99$. Lastly, we theoretically showed that our proposed convex formulation could be generalized to consider risk-constrained and asset allocation problems. Still, we left validating them through numerical experiments for future work.

\bibliographystyle{IEEEtran}
\bibliography{references}
\section*{Appendix}
\par \textit{Proof} of Theorem \ref{convex}: By Lemma \ref{lemma:R}, we only need to show that the reward function $R(\Theta_t)$ is concave for each $t$.
The Hessian $\mathbf{H}_{R_{t+1}}(\Theta_t)$ is given by:
\begin{equation}
    \begin{aligned}
    &\mathbf{H}_{R_{t+1}}(\Theta_t)=\\
    &\begin{bmatrix}[2]
        \frac{\partial^2R_{t+1}}{\partial\theta_1^2} & \frac{\partial^2R_{t+1}}{\partial\theta_1\partial\theta_2} & \cdots & \frac{\partial^2R_{t+1}}{\partial\theta_1\partial\theta_\kappa} & \frac{\partial^2R_{t+1}}{\partial\theta_1\partial\theta_{\kappa+1}}\\
        \frac{\partial^2R_{t+1}}{\partial\theta_2\partial\theta_1} & \frac{\partial^2R_{t+1}}{\partial\theta_2^2} & \cdots & \frac{\partial^2R_{t+1}}{\partial\theta_2\partial\theta_\kappa} & \frac{\partial^2R_{t+1}}{\partial\theta_2\partial\theta_{\kappa+1}}\\
        \vdots & \vdots & \ddots & \vdots & \vdots\\
        \frac{\partial^2R_{t+1}}{\partial\theta_\kappa \partial\theta_1} & \frac{\partial^2R_{t+1}}{\partial\theta_\kappa \partial\theta_2} & \cdots & \frac{\partial^2R_{t+1}}{\partial\theta_\kappa^2} & \frac{\partial^2R_{t+1}}{\partial\theta_\kappa \partial\theta_{\kappa+1}}\\
        \frac{\partial^2R_{t+1}}{\partial\theta_{\kappa+1}\partial\theta_1} & \frac{\partial^2R_{t+1}}{\partial\theta_{\kappa+1}\partial\theta_2} & \cdots & \frac{\partial^2R_{t+1}}{\partial\theta_{\kappa+1}\partial\theta_\kappa} & \frac{\partial^2R_{t+1}}{\partial\theta_{\kappa+1}^2}
    \end{bmatrix}
    \end{aligned}
    \label{eq:org_rewardHessian}
\end{equation}
Where the second partial derivatives in \eqref{eq:org_rewardHessian} are given by:
\begin{equation}
    \begin{aligned}
        \frac{\partial^2R_{t+1}}{\partial\theta_i\partial\theta_j}=
        \begin{cases}
            ax_{i,t}x_{j,t};      &\text{$i,j\leq\kappa\quad\mathrm{and}\quad i\neq j$}\\
            ax_{i,t}^2;     &\text{$i,j\leq\kappa\quad\mathrm{and}\quad i=j$}\\
            ax_{i,t}; &\text{$i\leq\kappa\quad\mathrm{and}\quad j=\kappa+1$}\\
            ax_{j,t}; &\text{$j\leq\kappa\quad\mathrm{and}\quad i=\kappa+1$}\\
            a;                   &\text{$i,j=\kappa+1$}
        \end{cases}&&
    \end{aligned}
    \label{eq:hessian_elements}
\end{equation}
For $i,j=1,2,...,\kappa+1$, and $a = \frac{d^2f}{dg_t^2}r_{t+1}$. It can be seen from \eqref{eq:hessian_elements} that the Hessian in \eqref{eq:org_rewardHessian} can be rewritten in terms of $a$ and a state's matrix $\mathbf{X}_t\in\mathbb{R}^{(\kappa+1)\times(\kappa+1)}$ as follows:
\begin{align}
    \notag
    &\mathbf{H}_{R_{t+1}}(\Theta_t)\equiv
    a\mathbf{X}_t=\\
    &\frac{d^2f}{dg_t^2}r_{t+1}
    \begin{bmatrix}[2]
        x_{1,t}^2 & x_{1,t}x_{2,t} & \cdots & x_{1,t}x_{\kappa,t} & x_{1,t}\\
        x_{2,t}x_{1,t} & x_{2,t}^2 & \cdots & x_{2,t}x_{\kappa,t} & x_{2,t}\\
        \vdots & \vdots & \ddots & \vdots &\vdots\\
        x_{\kappa,t}x_{1,t}& x_{\kappa,t}x_{2,t} & \cdots & x_{\kappa,t}^2&x_{\kappa,t}\\
        x_{1,t} & x_{2,t} & \cdots & x_{\kappa,t} & 1
    \end{bmatrix}
    \label{eq:rewardHessian}
\end{align}
We now prove that $\mathbf{X}_t$ is positive semi-definite by finding its eigenvalues denoted by $\lambda_i$ for $i=1, 2, ..., \kappa+1$. The characteristic polynomial of $\mathbf{X}_t$ is found to be:
\begin{equation}
    \label{eq:char_poly}
    \mathnormal{p}_{_{\mathbf{X}_t}}(\lambda) = \lambda^\kappa\bigg(\lambda-1-\sum_{n=1}^\kappa x_{n,t}^2 \bigg)
\end{equation}
From \eqref{eq:char_poly}, it can be seen that $\lambda_i=0$ for $i=1, 2, ..., \kappa$ while $\lambda_{\kappa+1} > 0$ since $\lambda_{\kappa+1}=1 + \sum_{n=1}^\kappa x_{n,t}^2$. Thus, the states' matrix $\mathbf{X}_t\geq0\quad\forall t\in T$. Then, the Hessian in \eqref{eq:rewardHessian} is negative semi-definite whenever \begin{equation}
    \frac{d^2f}{dg_t^2}r_{t+1} \leq 0
    \label{eq:concavity_condition}
\end{equation}
For each $t$, the condition \eqref{eq:concavity_condition} can be satisfied when one of the following is true:
\begin{flalign*}
    \text{1) $r_{t+1} \geq 0$ and $\frac{d^2f}{dg_t^2} \leq 0$ $\Longrightarrow$ $r_{t+1} \geq 0$ and $f$ is concave.}&&\\
    \text{2) $r_{t+1} \leq 0$ and $\frac{d^2f}{dg_t^2} \geq 0$ $\Longrightarrow$ $r_{t+1} \leq 0$ and $f$ is convex.}&&
\end{flalign*}

\par \textit{Proof} of Theorem \ref{convex_general2}: The Hessian of the reward in \eqref{eq:reward_portfolio} is given as:
\begin{equation}
    \mathbf{H}_{R^p_{t+1}}(\Theta_t) = \frac{d^2f}{dg_t^2}(r_{t+1} - r_{t+1}^f)\mathbf{X}_t
    \label{eq:hessian_portfolio}
\end{equation}
We can see that $\mathbf{H}_{R^p_{t+1}}$ is similar to $\mathbf{H}_{R_{t+1}}$ in \eqref{eq:rewardHessian}, except for the term $r_{t+1}$. Therefore, with replacing $r_{t+1}$ in \eqref{eq:rewardHessian} by $(r_{t+1} - r_{t+1}^f)$, the rest of the proof is analogous to the proof of Theorem \ref{convex}. 
\vspace{12pt}
\color{red}
\pdfoutput=1
\end{document}